\documentclass[aps,pra, twocolumn, showpacs]{revtex4}
\usepackage{amsfonts}
\usepackage{graphicx}
\usepackage{dcolumn}
\usepackage{bm}

\begin{document}

\preprint{preprint}
\title{Kinetic Thomas-Fermi solutions of the Gross-Pitaevskii equation}
\author{M. \"{O}lschl\"{a}ger$^{1}$, G. Wirth$^{1}$, C. Morais Smith$^{2}$, and A. Hemmerich$^{1}$}
\address{$^{1}$Institut f\"{u}r Laser-Physik, Universit\"{a}t Hamburg, 22761 Hamburg, Germany}
\address{$^{2}$Institute for Theoretical Physics, Utrecht University, 3508 TD Utrecht, The Netherlands}
\date{\today}

\begin{abstract}
Approximate solutions of the Gross-Pitaevskii (GP) equation, obtained upon neglection of the kinetic energy, are well known as Thomas-Fermi solutions. They are characterized by the compensation of the local potential by the collisional energy. In this article we consider exact solutions of the GP-equation with this property and definite values of the kinetic energy, which suggests the term "kinetic Thomas-Fermi" (KTF) solutions. We point out that a large class of light-shift potentials gives rise to KTF-solutions. As elementary examples, we consider one-dimensional and two-dimensional optical lattice scenarios, obtained by means of the superposition of two, three and four laser beams, and discuss the stability properties of the corresponding KTF-solutions. A general method is proposed to excite two-dimensional KTF-solutions in experiments by means of time-modulated light-shift potentials.
\end{abstract}


\maketitle

\section{Introduction}
\label{intro} 
A significant portion of the physics encountered in Bose-Einstein condensates (BECs) of ultra-cold atomic or molecular gases can be well described by a mean field approximation, which replaces the complicated many-body wave function by a single mean field obeying the famous Gross-Pitaevskii (GP) equation \cite{Pit:03}. This equation complements the Schr\"odinger equation by a non-linear term, which accounts for two-body interactions. In analogy to single particle physics, stationary states are determined by the time-independent GP-equation
\begin{equation}
\label{GP} 
\Big[\hat T  +  V( \mathbf{r})  + g |\psi( 
\mathbf{r})|^2\Big] \psi(\mathbf{r}) =\mu \,\psi(\mathbf{r}), 
\end{equation}
where $\hat T \equiv \frac{- \hbar^{2}}{2m}\Delta$ denotes the kinetic energy operator, $\mu$ is the chemical potential, $V( \mathbf{r})$ is the potential energy, and $g$ is a constant, which is determined by the binary collision cross-section in the $s$-wave approximation \cite{s-wave}. Not only the ground state of the system is often described with remarkable precision by means of the GP-equation but also the low energy excitation spectrum, including vortices and solitons  \cite{Pit:03, Dal:99}. Solutions $\psi$ of Eq.\ (\ref{GP}) may be characterized in terms of the following useful quantities: the local phase $S$ defined via $\psi =  |\psi| e^{i S}$, the particle density $\rho \equiv |\psi|^2$, the current density $\mathbf{j} \equiv (i \hbar /2m) (\psi \nabla \psi^{*} - \psi^{*} \nabla \psi)$, and the velocity field $\mathbf{v} \equiv \mathbf{j}/ \rho =  (\hbar /m) \nabla S$. Recall that $\nabla \cdot \mathbf{j} = 0$ and $\nabla \times \mathbf{v} = 0$ if $\rho \neq 0$, whereas the latter relation indicates that vortex filaments can only occur at density nodes.

In this paper we consider a class of excited stationary solutions of the GP-equation, which are characterized by a compensation of the local collisional and potential energies up to a spatially constant term, i.e., 
\begin{equation}
\label{KTF} 
V(\mathbf{r}) = V_{0}-g|\psi( \mathbf{r})|^2 \,\,. 
\end{equation}
This property is well known for approximate solutions derived in the so called Thomas-Fermi regime \cite{Pit:03}, where the kinetic energy in the GP-equation is neglected. Here, however, we are interested in exact solutions $\psi( \mathbf{r})$ of Eq.\ (\ref{GP}) subject to Eq.\ (\ref{KTF}), thus requiring that $\psi( \mathbf{r})$ is an eigenfunction of the kinetic energy operator, which suggests the term "kinetic Thomas-Fermi (KTF) solutions". KTF-solutions require some degree of non-linearity. They arise at the boundary between the weak interaction regime, where the dynamics retains Schr\"odinger character, and the strong interaction regime, where the inherent non-linearity of the GP-equation dominates. A natural environment for the emergence of KTF-solutions are optical lattices \cite{Hem:93, Gry:01, Lew:07} (i.e., ultracold gases subjected to periodic light-shift potentials), where they correspond to excited states at the edge of the first Brillouin zone. The interest in KTF-solutions in two- and three-dimensional optical lattices results from the combination of their remarkable formal simplicity with the possibility of non-zero current density fields, which can acquire unconventional spatial topologies,  for example, that of a vortex-antivortex sheet \cite{Hem:07}. This article presents a study of KTF-solutions in optical lattices, concerning their stability and their accessibility in experiments. In Sec. \ref{general} we point out that any light-shift potential arising in an arbitrary monochromatic light field with spatially constant polarization permits KTF-solutions of Eq.\ (\ref{GP}). In Sec. \ref{StabGen} we present a general discussion of the stability of KTF-solutions, which is applied to examples of particular interest for experiments in Sec. \ref{examples}. Here, we concentrate on one-dimensional (1D) and two-dimensional (2D) optical lattice scenarios readily obtained in experiments by means of the superposition of two, three and four laser beams. For the case of the 1D lattice we point out that the corresponding KTF-solution arises at the boundary between the regimes of linear Bloch bands and non-linear Bloch bands, characterized by additional loop structures \cite{Wu:00, Bro:01, Dia:02}. In Sec. \ref{excitation} we generalize the considerations of Ref. \cite{Hem:07} showing that for any 2D KTF-solution a suitable time-varying light-shift potential can be found, in order to drive the required current density. Finally, in Sec. \ref{bichromatic} this general scheme is applied to the examples of Sec. \ref{examples}.    

\section{KTF-solutions in arbitrary light-shift potentials}
\label{general}
KTF-solutions arise in a large class of light-shift potentials \cite{Let:68} according to the following scheme: Consider an arbitrary monochromatic light field $\mathbf{E}(\mathbf{r},t) \equiv (1/\sqrt{2}) (\textsf{\textbf{E}}(\mathbf{r}) e^{i \omega t} + \textsf{\textbf{E}}^{*}(\mathbf{r}) e^{-i \omega t})$ with $\textsf{\textbf{E}}(\mathbf{r}) \equiv \textsf{E}(\mathbf{r}) \,\hat \mathbf{\epsilon}$, where $\textsf{E}(\mathbf{r})$ is a complex scalar and $\hat \mathbf{\epsilon}$ is a spatially constant complex polarization vector satisfying $\hat \mathbf{\epsilon} \cdot \hat \mathbf{\epsilon}^{*} = 1$. Maxwells equations imply $(\Delta + k^2) \textsf{E}(\mathbf{r}) = 0$ with $k = \omega/ c$ and $\hat \mathbf{\epsilon} \cdot \nabla \textsf{E}(\mathbf{r}) = 0$. The light field $\mathbf{E}(\mathbf{r},t)$ gives rise to a light-shift potential $V( \mathbf{r}) \equiv - \Re(\alpha)\, \langle |\mathbf{E}(\mathbf{r},t)|^2\rangle = - \Re(\alpha)\,|\textsf{E}(\mathbf{r})|^2$, where $\alpha$ denotes the complex polarizability of the particles, which is assumed to be scalar, and the triangular brackets denote the time-average over one oscillation cycle. We now introduce the wave function $\psi(\mathbf{r}) \equiv \sqrt{\Re(\alpha)/g} \,\textsf{E}(\mathbf{r})$, which satisfies $\hat T \psi = E_{\rm rec} \psi$ with the single-photon recoil energy $E_{\rm rec} \equiv (\hbar k)^{2}/2m$. Moreover, Eq.\ (\ref{KTF}) holds with $V_{0}=0$ and thus Eq.\ (\ref{GP}) is satisfied with $\mu = E_{\rm rec}$. In brief, monochromatic light fields with spatially constant polarization give rise to KTF-solutions of the GP-equation with a particle density proportional to the time-averaged intensity. This assertion may be generalized to include a wider class of light fields admitting certain types of polarization gradients. Note that due to the vectorial character of the electric field, there is in general more than one light field yielding the same KTF-solution. For repulsive collisional interaction, $g > 0$ and thus $\Re(\alpha) > 0$ is required, which corresponds to normal dispersion, and thus a negative light shift potential, which is obtained for negative detuning of the light field with respect to the relevant atomic transition. In this case, the density maxima arise in the potential minima, i.e., the confining potential stabilizes the gas against collisional pressure. For attractive collisions and positive detuning, the density maxima coincide with the potential maxima, such that the repelling potential force counteracts collisional implosion of the gas.

\section{Stability analysis}
\label{StabGen}
Because KTF-solutions generally describe excitations, a central question in regard to their physical significance concerns their stability. The stability of solutions $\psi$ of Eq.\ (\ref{GP}) may be considered via the grand canonical potential $K[\psi_{\varepsilon}] \equiv \int d^{3}r [\psi_{\varepsilon}^{*}\hat T \psi_{\varepsilon} + (V-\mu) |\psi_{\varepsilon}|^2 + g |\psi_{\varepsilon}|^4/2 ]$ for $\psi_{\varepsilon} \equiv \psi + \varepsilon \chi$ with $\varepsilon \in \mathbb{R}$, and an arbitrary normalized wavefunction $\chi$. Use of Eq.\ (\ref{KTF}) and $\mu = E_{\rm rec}$ yields $\frac{\partial}{\partial \varepsilon} K[\psi_{\varepsilon}]_{\varepsilon=0} = 0$ and  
\begin{eqnarray}
\label{canonical} 
\frac{\partial^2}{\partial \varepsilon^2} K[\psi_{\varepsilon}]_{\varepsilon=0} &=& 
\nonumber \\ 
\int d^{3}r  [2 \chi^{*} (\hat T&-&E_{\rm rec}) \chi + g\,(\psi \chi^{*} + \psi^{*} \chi)^2 ], 
\end{eqnarray}
Stability requires that 
\begin{equation}
\label{stability} 
\frac{\partial^2}{\partial \varepsilon^2} K[\psi_{\varepsilon}]_{\varepsilon=0} > 0\,\,.
\end{equation}
Henceforth, we restrict ourselves to superpositions of $N$ optical travelling waves sharing the same polarization vector $\hat \mathbf{\epsilon}$ with arbitrary amplitudes and wave vectors $\mathbf{k}_{\nu}, \nu \in \{1,...,N\}$ with $k = |\mathbf{k}_{\nu}|$ for all $\nu$ and $\lambda =2\pi/k$ denoting the wavelength. The corresponding KTF-solution has the form $\psi \equiv \sum_{\nu = 1}^{N} \psi_{\nu} \, e^{i \mathbf{k}_{\nu} \mathbf{r}}$ with spatially constant complex amplitudes $\psi_{\nu}$. Different choices of $\mathbf{k}_{\nu}$ correspond to a rich variety of periodic and quasi-periodic light-shift potentials \cite{Gry:01}, including triangular, hexagonal or square lattice geometries. If the lattice potential is periodic, it suffices that each unit cell separately satisfies Eq.\ (\ref{stability}). In each unit cell we may then expand the arbitrary perturbation $\chi = \mathnormal{v}^{-3/2} \sum_{n_1,...,n_N \in \mathbb{Z}} C_{n_1,...,n_N} \prod_{\nu=1}^{N} e^{i n_{\nu} \mathbf{k}_{\nu} \mathbf{r}}$ in a Fourier-series with respect to the Bravais lattice, with $ \mathnormal{v}$ denoting the unit cell volume. Normalization requires $\sum_{n_1,...,n_N \in \mathbb{Z}} |C_{n_1,...,n_N}|^2 = 1$. For the kinetic term in Eq.\ (\ref{canonical}) we find $\int d^{3}r  2 \chi^{*} (\hat T - E_{\rm rec}) \chi$ = $2 E_{\rm rec} \sum_{n_1,...,n_N \in \mathbb{Z}} |C_{n_1,...,n_N}|^2 [(\sum_{\nu=1}^{N} n_{\nu} \hat \mathbf{k}_{\nu})^2 - 1]$ with $ \hat \mathbf{k}_{\nu} \equiv \mathbf{k}_{\nu}/k$. This term takes values larger than zero, if the expansion of $\chi$ comprises terms $C_{n_1,...,n_N}$ of sufficiently high order $\sum_{\nu=1}^{N} n_{\nu} \hat \mathbf{k}_{\nu}$, reflecting the fact that high frequency perturbations possess large kinetic energies. Let us assume repulsive collisions, i.e., $g > 0$. The collisional term $g (\psi \chi^{*} + \psi^{*} \chi)^2$ is positive then and $\frac{\partial^2}{\partial \varepsilon^2} K[\psi_{\varepsilon}]_{\varepsilon=0}$ becomes positive for perturbations with kinetic energies exceeding $E_{\rm rec}$, which thus do not contribute to possible instabilities of $\psi( \mathbf{r})$. If $\psi( \mathbf{r})$ comprises Fourier terms up to maximally first order, we may limit the Fourier expansion of relevant perturbations $\chi$ to second order. For attractive collisions ($g < 0$), the negative collisional term $g (\psi \chi^{*} + \psi^{*} \chi)^2$ does not permit stability. As discussed for specific examples below, stable KTF-solutions typically require $g \bar \rho > E_{\rm rec}$, where $\bar \rho$ denotes the mean particle density. Expressing $g \equiv 4 \pi \hbar^{2} a /m$ in terms of the $s$-wave scattering length $a$ and the atomic mass $m$, this leads to $a > a_{\rm min}$ with  $a_{\rm min} \equiv k^2 / 8\pi \bar \rho$. Inserting typical values ($k = 2 \pi \, 10^6$ m$^{-1}$, $\bar \rho = 10^{20}$ m$^{-3}$) yields $a_{\rm min} \approx 300\,a_0$ ($a_0 \equiv$ Bohr-radius), which is well in reach of experiments, if necessary by exploiting a Feshbach resonance \cite{ Pit:03}. 

\section{Examples of KTF-solutions in optical lattices}
\label{examples}
In the following, we apply the previous general considerations to three elementary examples: a 1D optical lattice and 2D optical lattices with triangular and square geometries. The collisional interaction is assumed to be repulsive and correspondingly the required light-shift potentials are negative. 

\subsection{1D Lattice} 
\begin{figure}
\includegraphics[scale=0.3]{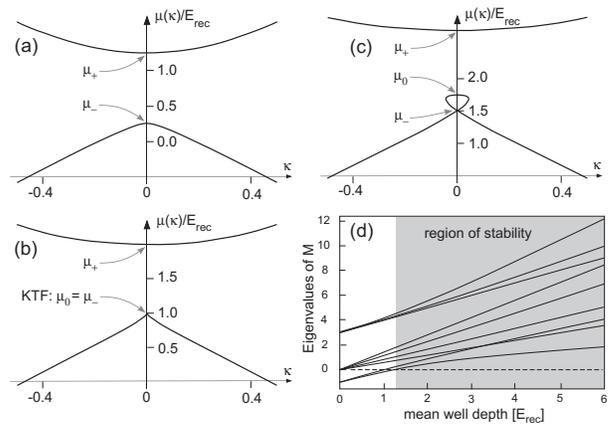}
\caption{\label{1D_Eigenvalues} The first and second Bloch bands derived from Eq.\ (\ref{loop}) are shown in (a), (b) and (c). (a) illustrates the linear regime ($\bar \rho = \frac{1}{2} \bar \rho_{c}$), (b) shows the boundary between the linear and the non-linear regime ($\bar \rho = \bar \rho_{c}$). The cusp at $\kappa = 0$ corresponds to the KTF solution $\phi(x) \equiv i \sqrt{2 \bar \rho_{c}} \sin(kx)$. (c) illustrates the non-linear regime ($\bar \rho = \frac{3}{2} \bar \rho_{c}$) characterized by a loop structure in the lower band. For all cases $\bar V = E_{\rm rec}$. In (d) the eigenvalues of $\mathbf{M}(z)$ for $N=2$ are plotted versus the mean well depth $z = \bar V / E_{\rm rec}$.}
\end{figure}

We begin with a 1D optical lattice composed of two counterpropagating travelling waves sharing the same polarization, thus yielding a light-shift potential $V(x) = - 2 \bar V \sin^2(kx)$ with the mean potential well depth $\bar V > 0$. The corresponding KTF-solution can be written as $\phi(x) \equiv i \sqrt{2 \bar \rho_{c}} \sin(kx)$, where in accordance with Eq.\ (\ref{KTF}) $\bar \rho_{c} \equiv \bar V/g$. The simplicity of this example lets us directly see that it marks the boundary between the regime of weak interaction, where the description in terms of Bloch states yields a conventional dispersion relation, and the strong interaction case, where the band structure acquires additional loops at the edges and the centre of the first Brillouin zone, with the consequence of inherently non-linear hysteretic dynamics \cite{Wu:00, Bro:01, Dia:02}. We suspect that this observation generalizes to be a generic property of KTF-solutions in optical lattices also in 2D and 3D. To clarify this point we remind ourselves (following Ref. \cite{Wu:00}) that the Bloch states near the edge of the first Brillouin zone may be approximated within a two-level picture as $\phi_{\alpha,\kappa}(x) = \sqrt{\bar \rho}\,(\cos{\alpha}\, e^{ik(1+\kappa)x}+\sin{\alpha}\, e^{ik(-1+\kappa)x})$ with $\kappa$ denoting the (small) deviation of the quasi-momentum from its value $k$ at the zone edge ($|\kappa| \ll 1$). The variational parameter $\alpha$ is determined by subjecting $\phi_{\alpha,\kappa}(x)$ to Eq.\ (\ref{GP}), which leads to the system of equations
\begin{eqnarray}
\label{loop} 
 \left(2 E_{\rm rec} + (3 \frac{\bar \rho}{\bar \rho_{c}} - 2)\, \bar V   - 2 \mu \right) \sin(2\alpha) + \bar V &=& 0 
\nonumber \\  \\  \nonumber
\left(4E_{\rm rec} \, \kappa  - \frac{\bar \rho}{\bar \rho_{c}} \bar V \cos(2\alpha) \right) \sin(2\alpha) - \bar V \cos(2\alpha) &=& 0   \,\, .
\end{eqnarray}
In Fig.~\ref{1D_Eigenvalues} we plot the chemical potential $\mu(\kappa)$, solution to Eq.\ (\ref{loop}), in the vicinity of the zone edge for the three cases $\bar \rho < \bar \rho_{c}$ (a), $\bar \rho = \bar \rho_{c}$ (b) and $\bar \rho > \bar \rho_{c}$ (c). The case $\bar \rho = \bar \rho_{c}$ (b), characterized by a cusp arising at the zone edge ($\kappa = 0$), separates the regimes with (c) and without (a) a loop structure in the dispersion function $\mu(\kappa)$. By setting $\kappa = 0$ in Eq.\ (\ref{loop})  one may directly determine the solutions at the zone boundary. Two solutions $\mu_{\pm} = E_{\rm rec} - \frac{1}{2} (2 - 3 \frac{\bar \rho}{\bar \rho_{c}} \mp 1) \bar V $ arise for arbitrary values of $\bar \rho$. If $\bar \rho \geq \bar \rho_{c}$ an additional solution (the top of the loop) $\mu_{0} = E_{\rm rec} - ( 1 - \frac{\bar \rho}{\bar \rho_{c}}) \bar V$ becomes possible, for which $\sin(2\alpha) = \frac{\bar \rho_{c}}{\bar \rho}$ holds. The corresponding wave function $\phi_{0}(x) \equiv i \sqrt{2 \bar \rho} \sin(kx)$ is identical with the KTF solution $\phi(x)$ at the critical density $\bar \rho = \bar \rho_{c}$ and the chemical potential becomes $\mu_{0} = E_{\rm rec}$ in this case. 

Let us next inspect the stability of the KTF-solution $\phi(x)$. Upon making a Fourier expansion of $\chi = \lambda^{-1/2}  \sum_{n \in \{-N,...,N\}} C_{n} e^{i n k x}$ up to $N$th order on the unit cell $[0, \lambda]$ we may
express $\frac{\partial^2}{\partial \varepsilon^2} K[\psi_{\varepsilon}]_{\varepsilon=0} = \mathbf{c} \, \mathbf{M}(z) \, \mathbf{c}$ as a bilinear form with respect to the vector $\mathbf{c} \equiv (\mathbf{a},\mathbf{b})$, where $\mathbf{a} \equiv \Re[(C_{-N},...,C_{N})]$, $\mathbf{b} \equiv \Im[(C_{-N},...,C_{N})]$ and
\begin{eqnarray}
\label{case1} 
\mathbf{M}(z) \equiv \left( \begin{array}{cc} \mathbf{M}^{(+)}(z) & 0  \\  0 & \mathbf{M}^{(-)}(z) \end{array}\right)
\end{eqnarray}
with $z = g \bar \rho_{c}/E_{\rm rec} = \bar V / E_{\rm rec}$ and the symmetric matrices $\mathbf{M}^{(\pm)}_{n,m}(z)$ $\equiv (n^2 + z - 1)\, \delta_{n,m} + \frac{1}{4} z \,(\delta_{n+2,m} + \delta_{n-2,m} + \delta_{n,m+2} + \delta_{n,m-2}) \pm \frac{1}{4} z \,(\delta_{n+2,-m} + \delta_{n-2,-m} + \delta_{-n,m+2} + \delta_{-n,m-2})$. Stability requires that all eigenvalues of $\mathbf{M}(z)$ exceed zero. Limiting the Fourier expansion to second order $N=2$ yields the eigenvalues shown in Fig.~\ref{1D_Eigenvalues}(d). The lowest eigenvalue crosses zero at $z = 1.22$, where $z E_{\rm rec}$ denotes the mean potential well depth. We also have extended the Fourier expansion up to third order obtaining the same result for the lowest lying branch of the eigenvalues in Fig.~\ref{1D_Eigenvalues}, thus confirming that in fact higher than second order terms in the expansion of $\chi$ are irrelevant for the stability.
\subsection{2D Square Lattice}
\begin{figure}
\includegraphics[scale=0.34]{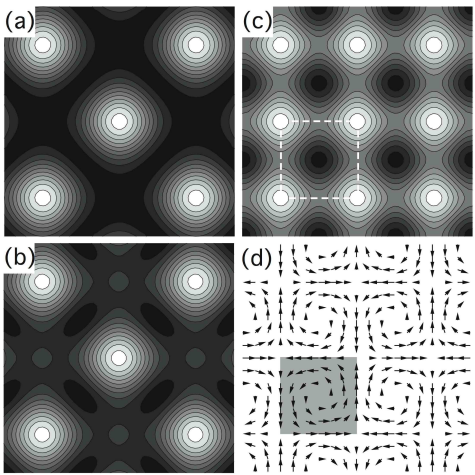}
\caption{\label{2D_Solutions} Local particle density $\rho_{\theta}(x,y)$ for $\theta = 0$ (a), $\theta = \pi/4$ (b), and $\theta = \pi/2$ (c). Black (white) indicates low (high) particle density.  In (d) the particle current density $\mathbf{j}_{\theta}(x,y)$ is shown for $\theta \not= n \pi$. The white dashed (c) and grey (d) rectangles shows a $\lambda/2 \times \lambda/2$ sized plaquette. In all graphs an area corresponding to $3\times3$ plaquettes are shown.}
\end{figure}

\begin{figure}
\includegraphics[scale=0.24]{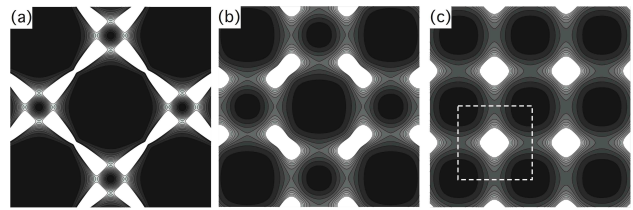}
\caption{\label{Ekin} The square of the velocity field $\mathbf{v}_{\theta}^2$ is plotted for $\theta = \pi/10$ (a), $\theta = \pi/4$ (b) and $\theta = \pi/2$ (c). The same areas as in Fig.~\ref{2D_Solutions} are shown.}
\end{figure}

Next, we discuss a 2D optical square lattice scenario composed of two optical standing waves $\hat \mathbf{z} \sin(kx)$ and $\hat \mathbf{z} \sin(ky)$), oriented along the $x$- and $y$-axes with linear polarizations parallel to the $z$-axis. We admit an arbitrary phase lag $\theta$ between the oscillations of the two standing waves. The specific case $\theta = \pi/2$ has been discussed in Ref. \cite{Hem:07}. Apart from the interesting additional physics accessible, the inclusion of other values of $\theta$ in our analysis is essential for experiments, because $\theta$ can only be controlled with finite precision. Light-field configurations of this type have been extensively applied in numerous previous experiments \cite{Hem:91}. The corresponding KTF-solution is $\psi_{\theta}(x,y) \equiv \sqrt{\bar \rho} \, (e^{i \theta/2} \sin(kx) + e^{-i \theta/2} \sin(ky))$ with $\bar \rho$ denoting the mean particle density. According to Eq.\ (\ref{KTF}) the required light-shift potential is $V_{\theta}(x,y) = - \bar V (\sin^2(kx) + \sin^2(ky) + 2 \cos(\theta) \sin(kx) \sin(ky))$ with $\bar V = g \bar \rho$. Despite its formal simplicity, $\psi_{\theta}(x,y)$ possesses remarkable properties. For $\theta = n \pi$ with integer $n$, $\psi_{\theta}(x,y)$ represents a two-dimensional array of stationary solitons separated by nodal lines, where the particle density $\rho_{\theta}(x,y)$ becomes zero (black regions in Fig.~\ref{2D_Solutions}(a)). For values of $\theta \neq n \pi$ the particle density nodes are points (cf. Fig.~\ref{2D_Solutions} (b), (c)) and a periodic pattern of vortical fluxes arises with alternating rotational sense for adjacent plaquettes. This is illustrated in Fig.~\ref{2D_Solutions} (d), which shows the particle flux density for $\theta \not= n \pi$, given by $\mathbf{j}_{\theta}(x,y) = (\hbar/ m) \, \bar \rho  \, \sin(\theta) \, \mathbf{\nabla} \times \mathbf{\hat z}  \, \sin(kx) \sin(ky)$. While $\mathbf{j}_{\theta}$ scales with $\sin(\theta)$, its spatial structure is the same for all $\theta$. In contrast, the spatial structure of  the corresponding velocity field $\mathbf{v}_{\theta}$ depends on $\theta$. This is shown in Fig.~\ref{Ekin}, where $\mathbf{v}_{\theta}^2$ is plotted for $\theta = \pi/10$ (a), $\theta = \pi/4$ (b) and $\theta = \pi/2$ (c). If $\theta$ tends to zero, each contour for some fixed value of $|\mathbf{v}_{\theta}|$ approaches the diagonal nodal line structure of the $\theta = 0$ density distribution (cf. Fig.~\ref{2D_Solutions} (a)). Nevertheless, for all $\theta \neq n \pi$, one obtains 
\begin{eqnarray}
\label{vorticity} 
\mathbf{\nabla} \times \mathbf{v}_{\theta}(x,y) =
\\  \nonumber
\mathbf{\hat z}\,\, 2 \pi \,\, \frac{ \hbar}{m} \sum_{n,m \in \mathbb{Z}} &(-1)^{n+m}& \,\, \delta \left(x-\frac{n\pi}{k},y-\frac{m\pi}{k}\right),
\end{eqnarray}
showing that $\mathbf{v}_{\theta}(x,y)$ represents a pure vortex-anti-vortex lattice with vortex-filaments at positions $kx,ky \in \pi \, \mathbb{Z}$. For $g > 0$ the vortices are pinned at the potential maxima, which correspond to the density minima of the lattice (black regions in Fig.~\ref{2D_Solutions} (b),(c)), in accordance with results obtained for plain vortex lattices prepared in large scale traps and subsequently exposed to an optical lattice potential \cite{Rei:05, Tun:06}. 

For a superfluid, we expect the particle density to vanish at the poles of the velocity field on a spatial scale determined by the healing length. We briefly discuss the implications for the vortex cores of the $\theta = \pi/2$-case, which have a radius on the order of $R_{\rm core} \approx 1/k$. Defining the healing length as $\xi \equiv (8\pi a \bar \rho)^{-1/2}$ with $a$ denoting the $s$-wave scattering length, upon use of Eq.\ (\ref{KTF}) one may write $(k \xi)^2 = E_{\rm rec}/g \bar \rho= E_{\rm rec}/ \bar V$ with mean potential well depth $\bar V$. Thus, the condition that the core size exceeds the healing length $\xi < R_{\rm core}$ is equivalent to $E_{\rm rec} <  \bar V$. 

The mean kinetic energy per particle $T_{\theta} \equiv \int d^{3}r\, \rho_{\theta} \frac{m}{2}\mathbf{v}_{\theta}^2/ \int d^{3}r \rho_{\theta}$ connected with the velocity field $\mathbf{v}_{\theta}$ increases from zero to its maximal value $\frac{2}{\pi} E_{\rm rec}$ as $\theta$ is tuned from zero to $\pi/2$. The total kinetic energy per particle is $E_{\rm rec}$ independent of $\theta$. The difference $E_{\rm rec} - T_{\theta}$ reflects the quantum pressure, corresponding to the $\theta$-dependent degree of localization. The angular momentum per particle is $\mathbf{L}_{\theta} \equiv \int d^{3}r \,\mathbf{r} \times \mathbf{j}_{\theta}/ \int d^{3}r\,\rho_{\theta} = (8/\pi^2) \, \hbar\,\sin(\theta)\, \hat \mathbf{z}$. The small difference between the factor $8/\pi^2$ and unity accounts for the fact that the velocity fields from different vortices with opposite sense of rotation yield some cancellation.

\begin{figure}
\includegraphics[scale=0.6]{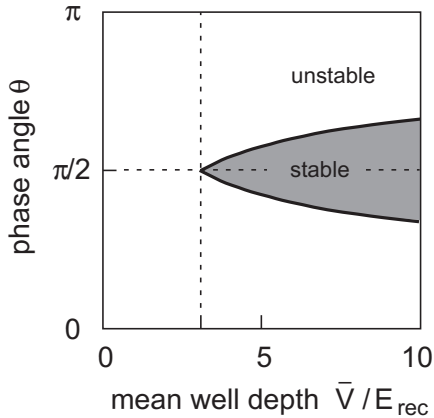}
\caption{\label{2D_Stability} Stability diagram for the family of KTF-solutions $\psi_{\theta}(x,y) \equiv \sqrt{\bar \rho} \, (e^{i \theta/2} \sin(kx) + e^{-i \theta/2} \sin(ky))$ with mean particle density $\bar \rho$ and light-shift potential $V_{\theta}(x,y) = - \bar V\,(\sin^2(kx) + \sin^2(ky) + 2 \cos(\theta) \sin(kx) \sin(ky))$.}
\end{figure}

We may consider the stability of $\psi_{\theta}(x,y)$ following the same procedure as in the 1D case illustrated in the context of Eq.\ (\ref{case1}). Here, a Fourier expansion up to second order leads to a $26\times26$ matrix, which is diagonalized in order to obtain the ($g > 0$) stability diagram shown in Fig.~\ref{2D_Stability}. As is seen in the figure, the vortical particle flux $\mathbf{j}_{\theta}$ tends to stabilize $\psi_{\theta}(x,y)$. The largest stability range ($\bar V/E_{\rm rec} \gtrsim 3.1$ ) arises for $\theta = (n+\frac{1}{2}) \pi$ corresponding to Fig.~\ref{2D_Solutions} (c) and Fig.~\ref{Ekin} (c). For $\theta = n \pi$ the flux vanishes and instability occurs for any potential well depth. For negative $g$-values, $\psi_{\theta}(x,y)$ becomes unstable for all $\theta$.

\subsection{2D Triangular Lattice}
\begin{figure}
\includegraphics[scale=0.3]{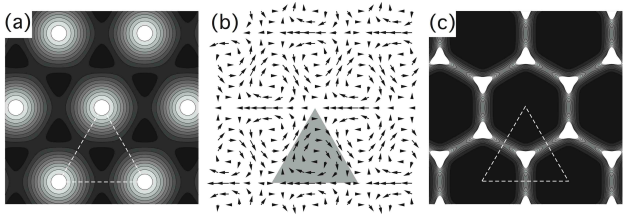}
\caption{\label{2D_Triangular} The particle density $\rho_{\triangleleft}(x,y)$ (a), the flux density $\mathbf{j}_{\triangleleft}(x,y)$ (b), and the square of the velocity field $\mathbf{v}_{\triangleleft}^2(x,y)$ (c) are plotted. Black (white) indicates low (high) values. The equilateral triangles (with $\lambda/2$ side length) indicate a single plaquette of the lattice.}
\end{figure}

As a third example, we briefly discuss a triangular lattice composed of three travelling waves propagating within the $xy$-plane with linear polarizations parallel to the $z$-axis and wave-vectors $\mathbf{k}_{\nu} = k\,\{\cos(2\pi\nu/3),\sin(2\pi\nu/3)\},\nu\in\{1,2,3\}$, mutually enclosing a $120^{\circ}$ angle. The corresponding KTF-solution is $\psi_{\triangleleft}(x,y) \equiv \sqrt{\bar \rho/3} \, (e^{i kx} + e^{-i kx/2} \cos(\frac{\sqrt{3}}{2}ky))$. In contrast to the superposition of four travelling waves, there is no free phase parameter here. Inserting complex amplitudes for the three superimposed travelling waves would merely yield a spatial shift of the resulting field \cite{Gry:01}. In Fig.~\ref{2D_Triangular} we show the particle density $\rho_{\triangleleft}(x,y)$ (a), the flux density $\mathbf{j}_{\triangleleft}(x,y)$ (b) and the square of the velocity field $\mathbf{v}_{\triangleleft}^2(x,y)$ (c). We encounter a situation very similar to the case of the square lattice at $\theta = \pi/2$ discussed in Fig.~\ref{2D_Solutions} (c), (d) and Fig.~\ref{Ekin} (c), obtaining a vortex-anti-vortex lattice with vortex-filaments at the nodes of the particle density.

\section{Excitation of KTF-solutions}
\label{excitation}
In order to study KTF-solutions in experiments, a technique is required to excite them. A well-known method to produce excitations with a definite momentum in BECs is stimulated Raman scattering (SRS). Two intersecting laser beams with common polarizations, frequencies $\omega$ and $\omega+\Omega$, and $k$-vectors $\mathbf{k}_1, \mathbf{k}_2$ are employed to produce a travelling intensity grating moving at a speed $c_{\rm{g}} \equiv \Omega / \Delta k$ with $\Delta k \equiv |\mathbf{k}_1 - \mathbf{k}_2|$. This light grating yields a corresponding moving light-shift potential, which can excite a velocity field, if $\Omega$ and $\Delta k$ match with the Bogoliubov dispersion relation \cite{Koz:99, Ste:99}, which requires $c_{\rm{g}} = \sqrt{(\hbar \Delta k /2m)^2 + c_{\rm{s}} }$ with the sound velocity $c_{\rm{s}} \equiv \sqrt{\bar \rho g /m}$. In the regime of phonon-like excitations ($(\hbar \Delta k)^2/2m \ll \bar \rho g$), the resonance condition is approximated by $c_{\rm{g}} = c_{\rm{s}}$. SRS can be extended to yield excitations with more complex spatial geometries. For example, the KTF-solution $\psi_{\theta}(x,y)$ comprises the four components $\pm \hbar k \, \hat \mathbf{x}, \pm \hbar k \, \hat \mathbf{y}$ in momentum space, where $\hat \mathbf{x}, \hat \mathbf{y}$ denote the unit vectors in $x$- and $y$-directions. The excitation of each momentum component requires a pair of counterpropagating laser beams with frequencies $\omega$ and $\omega+\Omega$, i.e., in total eight beams are necessary forming a pair of crossed bichromatic standing waves. Similar multi-beam variants of SRS have been proposed as a means to excite vortex or skyrmion states in a BEC \cite{VortexExcitation, SkyrmionExcitation}. 

In the following, we discuss the use of SRS to excite arbitrary 2D KTF-solutions $\psi(x,y) = |\psi(x,y)|\, e^{i S(x,y)}$ in the $xy$-plane, which may possess velocity fields with a complex geometry. Since $\psi(x,y)$ solves the Helmholtz equation we may consider the monochromatic light-field $\mathbf{E}_1(x,y,t) \equiv (1/\sqrt{2}) [\textsf{\textbf{E}}(x,y) e^{i \omega t} + \textsf{\textbf{E}}^{*}(x,y) e^{-i \omega t}]$ with $\textsf{\textbf{E}}(x,y) \equiv \sqrt{g/ \Re(\alpha)} \,\psi(x,y)\, \hat \mathbf{z}$ and the corresponding light-shift potential $V(x,y) =$ $-\Re(\alpha) |\textsf{\textbf{E}}(x,y)|^2 = - g |\psi(x,y)|^2$. Consider the following bichromatic extension $\mathbf{E}_2(x,y,t) \equiv (1/\sqrt{2}) [\textsf{\textbf{E}}^{\Omega}(x,y,t) e^{i \omega t} + \textsf{\textbf{E}}^{\Omega*}(x,y,t) e^{-i \omega t}]$ with $\textsf{\textbf{E}}^{\Omega}(x,y,t) \equiv \cos(\eta)\, \textsf{\textbf{E}}(x,y) + \sin(\eta)\, e^{i \Omega t}\, \textsf{\textbf{E}}^{*}(x,y)$ (with some $\eta \in [0,\pi/2]$). Since $\mathbf{E}_1(x,y,t)$ by definition solves Maxwells equation, this holds for $\mathbf{E}_2(x,y,t)$ in very good approximation, if $\Omega \ll \omega$ is assumed. Note that in the following examples $\Omega/\omega$ is on the order of $10^{-11}$. The bichromatic field $\mathbf{E}_2(x,y,t)$ yields the light-shift potential $V^{\Omega}(x,y,t) = -\Re(\alpha) |\textsf{\textbf{E}}^{\Omega}(x,y,t)|^2 = V(x,y) \left(1 +  \sin(2\eta) \cos \left(2S(x,y) - \Omega t \right) \right)$. Hence, $V^{\Omega}(x,y,t)$ is a sum of the stationary light-shift potential $V(x,y)$ satisfying Eq.\ (\ref{KTF}) and a modulation term $V^{\rm{mod}}(x,y,t) \equiv \sin(2\eta) V(x,y) \cos\left(2S(x,y) - \Omega t \right)$ with an experimentally adjustable modulation strength $\sin(2\eta)$. $V^{\rm{mod}}(x,y,t)$ provides a light-shift grating moving according to the wave-vector field $\mathbf{K}(x,y) \equiv 2 \nabla S(x,y)$ directly proportional to the velocity field $\mathbf{v}(x,y)$ of $\psi(x,y)$. Thus, temporary application of $V^{\rm{mod}}(x,y,t)$ should be a means to excite the KTF-solution $\psi(x,y)$, if $\hbar \Omega$ is adjusted to match the energy difference between $\psi(x,y)$ and the ground state in the lattice potential $V(x,y)$ \cite{Yuk:02}. Since we are interested in solutions $\psi(x,y)$, for which the collisional energy $g \bar \rho$ is on the same order as the kinetic energy, we cannot directly apply the theory of Ref. \cite{Yuk:02} to calculate the excitation efficiency. An extension of Ref. \cite{Yuk:02} to the regime of significant collisional interaction is a difficult venture, which requires further research. In case of a stable KTF-solution, the fact that we can provide the appropriate time-dependent potential, which drives the required current density, is nevertheless a strong indication, that this KTF-solution can be efficiently excited. 

\section{Examples of bichromatic light-shift potentials}
\label{bichromatic}
\begin{figure}
\includegraphics[scale=0.28]{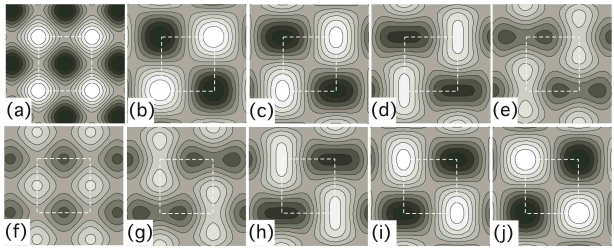}
\caption{\label{Rotor} In (a) $V_{\pi/2}(x,y)$ is replotted from Fig.~\ref{2D_Solutions}(c). White (black) regions indicate minima (maxima). In (b)-(j) the micro-rotor potential $V^{\rm{mod}}_{\pi/2}(x,y,t)$ is shown for $\Omega t = n \pi/8$ with $n \in \{0,1,...,8\}$. The same area as in (a) is shown. The dashed white squares indicate a $\lambda/2\times \lambda/2$-sized plaquette.}
\end{figure}
In the following, we illustrate the SRS-method in case of the example $\psi_{\theta}(x,y) = |\psi_{\theta}(x,y)| \,e^{i S_{\theta}(x,y)}$ introduced in the second paragraph of Sec. \ref{examples}. The required bichromatic light field is $\textsf{\textbf{E}}^{\Omega}_{\theta}(x,y,t) \equiv \cos(\eta)\, \textsf{\textbf{E}}_{\theta}(x,y) + \sin(\eta)\, e^{i \Omega t}\, \textsf{\textbf{E}}_{\theta}^{*}(x,y)$ with $\textsf{\textbf{E}}_{\theta}(x,y) \equiv \psi_{\theta}(x,y) \sqrt{g/ \Re(\alpha)} \, \hat \mathbf{z}$. The corresponding bichromatic light-shift potential becomes  $V^{\Omega}_{\theta}(x,y,t) =   V_{\theta}(x,y) + V^{\rm{mod}}_{\theta}(x,y,t)$ with $V^{\rm{mod}}_{\theta}(x,y,t) = V_{\theta}(x,y) \sin(2\eta) \cos \left(2S_{\theta}(x,y) - \Omega t \right)$. The time-evolution of the modulation $V^{\rm{mod}}_{\theta}(x,y,t)$ is illustrated in Fig.~\ref{Rotor} for $\theta = \pi/2$, indicating that it acts as a collection of microscopic rotors, which apply angular momentum with alternating sign within the plaquettes of the square lattice $V_{\pi/2}(x,y)$. In the vicinity of each maximum of $V_{\pi/2}(x,y)$ (replotted in (a) from Fig.~\ref{2D_Solutions}(c)) the micro-rotor term $V^{\rm{mod}}_{\pi/2}(x,y,t)$ provides a quadrupole potential rotating with alternating directions for adjacent plaquettes. In (b)-(j) $V^{\rm{mod}}_{\pi/2}(x,y,t)$ is shown for $\Omega t = n \pi/8$, with $n \in \{0,1,...,8\}$, thus illustrating a $\Omega_{R} \, t = \pi/2$ clockwise rotation with angular frequency $\Omega_{R} = \Omega /2$. Let us briefly estimate the resonance condition. The angular momentum applied to each plaquette is approximately given by $m \, \Omega_R \, r^2$, where $r \equiv \lambda/4$ is the distance from the centre to the edge of the plaquette. Excitation of vortices requires an angular momentum of $\hbar$ per plaquette, i.e., $m \, \Omega_R \, r^2 \approx \hbar$ and thus $\hbar  \, \Omega \approx (8/ \pi^2) \, 2E_{\rm rec}$. For rubidium atoms and a convenient lattice wavelength ($\lambda$ = 1030 nm) $\Omega/2\pi = 3.5$ kHz. It has been recently pointed out that within a description of the optical lattice with the periodic potential $V_{\pi/2}(x,y)$ in terms of a Bose-Hubbard model \cite{BH}, the rotor potential $V^{\rm{mod}}_{\pi/2}(x,y,t)$ simulates a staggered magnetic field alternating for adjacent plaquettes  \cite{Lim:08}. Finally, we note that the modulation term $V^{\rm{mod}}_{\triangleleft}(x,y,t)$ obtained for the triangular KTF-solution $\psi_{\triangleleft}(x,y)$ of Fig.~\ref{2D_Triangular} (cf. Sec. \ref{examples}) displays a similar behavior as that observed in Fig.~\ref{Rotor} for the square lattice: Centered at each density node, microscopic rotors with quadrupolar shape apply angular momentum to the triangular plaquettes of Fig.~\ref{2D_Triangular}.

\begin{figure}
\includegraphics[scale=0.3]{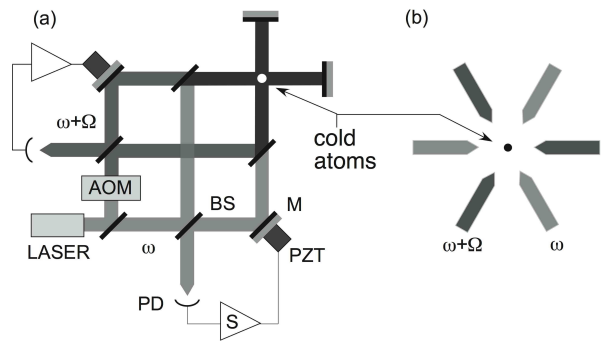}
\caption{\label{Michelson} (a) Optical set-up consisting of two nested Michelson-interferometers. PZT = piezo-electric transducer, M = mirror, BS = beam splitter, AOM = acousto-optic frequency modulator, PD = photo diode, S = servo electronics. (b) Beam configuration for triangular lattice.}
\end{figure}
Experimentally, the generation of the bichromatic light field $\textsf{\textbf{E}}^{\Omega}_{\theta}(x,y,t)$, appropriate for the excitation of $\psi_{\theta}(x,y)$, is straightforward using the optical set-up illustrated in Fig.~\ref{Michelson} (a), thus extending a method proven practicable in previous experiments \cite{Hem:91}. The monochromatic components $\textsf{\textbf{E}}_{\theta}(x,y)$ and $e^{i \Omega t}\, \textsf{\textbf{E}}_{\theta}^{*}(x,y)$ are produced in two nested Michelson-interferometers, each with its two branches folded under 90$^\circ$ angle. Two laser beams with adjustable amplitudes, frequency separation and linear polarizations perpendicular to the drawing plane in Fig.~\ref{Michelson} are used to couple both interferometers, which comprise piezo-electrically driven mirrors for servo control of the optical path length differences and thus the  temporal phase differences between the standing waves produced in each interferometer branch. The requirement, that the spatial fields in the two interferometers are complex conjugates, is realized by setting their  temporal phase differences to $\theta$ and $-\theta$, respectively. This corresponds to the adjustment of optical path length differences $\Delta l = \pm\,\theta \lambda / 2 \pi$. The frequency difference $\Omega$ and the modulation strength $\sin(2 \eta)$ are controlled by the acousto-optic frequency modulator shown in Fig.~\ref{Michelson} (a). 

Providing the bichromatic light-field for excitation of the triangular KTF-solution $\psi_{\triangleleft}(x,y)$ requires the superposition of six light beams with frequencies $\omega$ and $\omega+\Omega$ according to the sketch in Fig.~\ref{Michelson} (b). All beams share the same linear polarization perpendicular to the drawing plane in Fig.~\ref{Michelson} (b). Three beams with the same frequency $\omega$ are used to produce the light field $\textsf{\textbf{E}}_{\triangleleft}(x,y) = \sqrt{g/ \Re(\alpha)}\, \psi_{\triangleleft}(x,y)\, \hat \mathbf{z}$. The phase differences between these beams determine the position of the nodes of $\textsf{\textbf{E}}_{\triangleleft}(x,y)$. To ensure that the three beams at frequency $\omega+\Omega$ yield the corresponding field $e^{i \Omega t}\, \textsf{\textbf{E}}^{*}(x,y)$, their phase differences need to be the same as those between the $\omega$-beams. Unfortunately, because all beams with the same frequency travel along different paths, an interferometric realization of this condition is not easily implemented.

\section{Conclusions}
\label{concs}
In summary, we have introduced a family of stationary solutions of the Gross-Pitaevskii equation with definite values of the kinetic energy, for which the local collisional energy is compensated by the potential energy. In view of the fact that this property is shared with the well known approximate Thomas-Fermi solutions, obtained upon neglection of kinetic energy, we have suggested the term "kinetic Thomas-Fermi" (KTF) solutions. Such solutions are particularly relevant in the context of optical lattice scenarios, where they represent excited states at the edge of the first Brillouin zone. Conditions for the stability of KTF-solutions have been discussed and a general method has been proposed to excite KTF-solutions in experiments by means of time-modulated light-shift potentials. We have applied our general considerations to a few elementary examples: a 1D optical lattice, a 2D square lattice and a 2D triangular lattice, however, they should apply to more complex lattice geometries including quasi-periodic lattices. 

\section{Acknowledgments}
\label{acks}
We thank Lih-King Lim and Claus Zimmermann for helpful discussions. AH acknowledges support by DFG (He2334/10-1).

\end{document}